# Explaining Emergence[1]

HERVE ZWIRN

*Centre Borelli (ENS Paris Saclay, France) & IHPST (CNRS & University Paris 1, France)*

___________________________________

**ABSTRACT** Emergence is a pregnant property in various fields. It is the fact for a phenomenon to appear surprisingly and to be such that it seems at first sight that it is not possible to predict its apparition. That is the reason why it has often been said that emergence is a subjective property relative to the observer. Some mathematical systems having very simple and deterministic rules nevertheless show emergent behavior. Studying these systems shed a new light on the subject and allows to define a new concept, computational irreducibility, which deals with behaviors that even though they are totally deterministic cannot be predicted without simulating them. Computational irreducibility is then a key for understanding emergent phenomena from an objective point of view that does not need the mention of any observer.

## 1. Introduction

Emerging phenomena are present in almost all fields. They can be found in mathematics, physics, biology, economics, sociology and even painting. It is thus common to say that life or consciousness are emerging phenomena, astronomers can describe the shape of Saturn's rings as emerging, economists consider the currency as an emerging institution, the formation of a vortex is for physicists an emerging property during the flow of a fluid, the majority opinion that emerges from an exchange between many individuals of a population is emerging, the sudden appearance of a traffic jam depending on the behaviour of motorists or the image that suddenly appears to us when we step back looking at a pointillist painting are also described as emerging. We could multiply the examples by borrowing them from all areas of the world around us[2]. The systems showing this type of properties are mostly systems that are now called complex systems and whose holistic behaviour results from interactions between all the components that make them up[3].

Used loosely, the term "emergent" is attributed to phenomena of very diverse natures, but whose common characteristic is that they are in some way unforeseen or surprising. In any case, these phenomena are characterized by manifestations that are neither easy to explain nor to understand, even when we know a priori the elements that create them. Over and above the different forms these phenomena take, they share the fact that if we take an a priori view of only the elementary components of the system that make them up and the laws that govern them, we will have a hard time predicting their appearance. It's only by observing the dynamics of each system that we will be able to see what appears, which in these specific cases will take on a form we probably wouldn't have anticipated. In this case, we speak of the irreducibility of the global level (corresponding to the scale at which we observe the emerging phenomenon) to the local level (corresponding to the scale of the constituents). This simply means that it is at best extremely difficult, and at worst totally impossible, to predict the behaviour of the global level on the basis of knowledge of the local level.

---

[1] This paper will appear in the forthcoming proceedings of the UM6P Science Week 2023 – Complexity Summit.

[2] B. Walliser, H. Bersini, G. Lecointre, H. Zwirn « L'émergence, un concept pour comprendre les transitions » *in* M. Gargaud et G. Lecointre (eds.), *L'évolution de l'univers aux sociétés*, Editions Matériologiques, 2015.

[3] H. Zwirn, *Les systèmes complexes*, Odile Jacob, 2006.

This aspect of surprise, which is somehow linked to us, has led some to say that emergence is a subjective property, dependent on the observer, and that it is not possible to give it an objective definition. In this view, emergence is only in the eye of the observer and has no reality of its own. At the opposite extreme, the British emergentists of the late 19th and early 20th centuries defended a thesis known as strong emergence, according to which, for certain configurations of a system's components, new forces appear: for them, these forces are in no way reducible to the forces governing the behaviour of the components themselves; they are totally independent of them and result from the configuration adopted by the constituents. In this sense, the irreducibility of the global level to the local level becomes essential, and it is therefore inherently impossible to predict global behaviour from knowledge of the local level. Of course, nobody defends such a position these days.

Here, I would like to present a conception of emergence that is objective in the sense that it can be defined without reference to an observer, and yet in no way requires the appearance of new and different forces at the global level, as postulated by British emergentists. This conception calls on the concept of computational irreducibility, which I shall begin by describing. To do so, it is useful to start with a few illustrative examples of this type of operation.

## 2. Simple systems whose behaviour is hard to predict

As we have just said, emergence is a property that applies to a wide range of fields, so the analysis of emergent phenomena is often made difficult by the fact that these phenomena are not isolated and clearly individualized, but occur within a complex environment. The reasons why they are emergent are then often drowned out by ancillary effects, which have nothing to do with the emergence itself but are inseparable from the phenomenon itself. This is the case in disciplines such as biology, economics or sociology, where the description and identification of the phenomenon itself is already non-trivial, and where, what's more, the causes at the origin of the phenomenon are often poorly known (whether we are thinking of life or the emergence of a majority opinion on a cultural subject). For this reason, as our aim is to lay bare the very essence of what makes a phenomenon emergent, we will be focusing on an extremely precise domain, that of certain mathematical objects, capable of producing quite spectacular emergent effects, but which have the advantage that it is possible to describe very precisely the components of which they are made and the rules of behaviour of these components. Freed from irrelevant details, it then becomes possible to explain the essence of the emergence mechanism, which also applies to other phenomena but in a less directly accessible way.

### 2.1 Mono dimensional cellular automata
Cellular automata were invented by John von Neumann and Stanislas Ulam in the 1940s. A cellular automaton is a dynamic system consisting of a grid of cells in a one- or multi-dimensional array. Each cell contains a value taken from a finite set of possibilities. The system evolves over time, following rules for updating each cell according to the values contained in neighbouring cells. We shall start with the simplest of these, the two-colour, one-dimensional nearest-neighbour automata. Each cell can have two values (0 or 1, or be white or black), and the update rules depend only on the colour of the cell concerned and that of its two nearest neighbours. Evolution takes place in discrete time steps, and successive stages are represented as rows of cells stacked one below the other. Wolfram proposed a numbering system to designate each of the 256 cellular automata of this type, and we will use this numbering system, which is used by everyone[4]. It works as follows. Each group of 3 cells can be in one of the 8 possible cases below (see figure 1):

---

[4] S. Wolfram, *A new kind of science*, Champaign, IL: Wolfram Media, Inc., 2002.

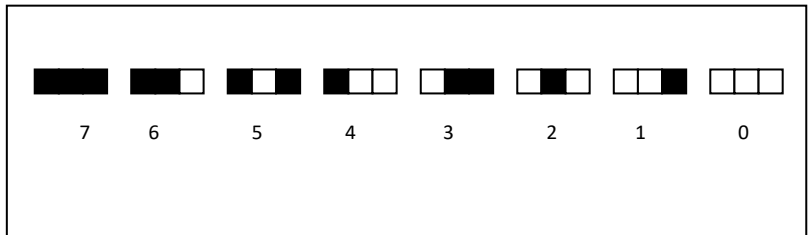

*Figure 1. The different configurations of 3 cells*

For each of these configurations, an automaton rule specifies the colour of the middle square in the next step. The rule shown in figure 2 is rule no. 254, for the reason explained below.

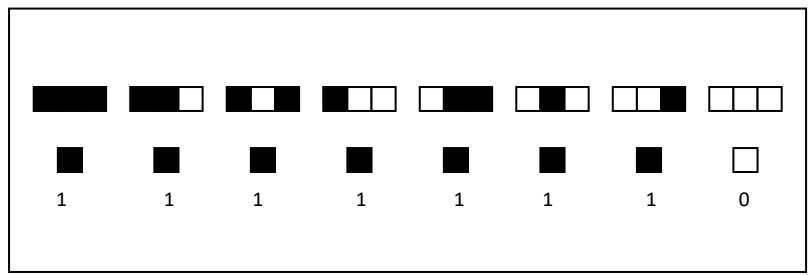

*Figure 2. Automaton number 254*

This rule states that if the top three squares are white, then the middle square will be white in the next step; if the right-hand square is black and the other two are white, then the middle square will be black in the next step, and so on. By associating the value 1 with the black squares and 0 with the white squares, and writing the rule as in figure 2 (i.e., ordering the configurations of three squares by the value they represent in binary), we see that we obtain another binary number given by the value of the squares in the second line. This number is the rule number. Here, "11111110" is the binary value of 254.

Figure 3 shows some of the steps involved in automaton no. 254:

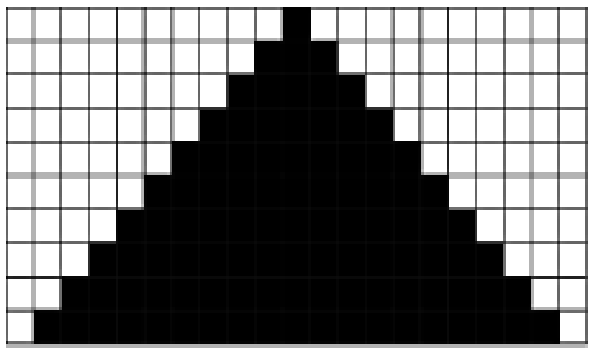

*Figure 3. Automaton no.254*

This automaton is not very interesting since it simply fills the plan with black boxes. Figure 4 shows a more interesting behaviour, although simple and regular, that of automata no. 90:

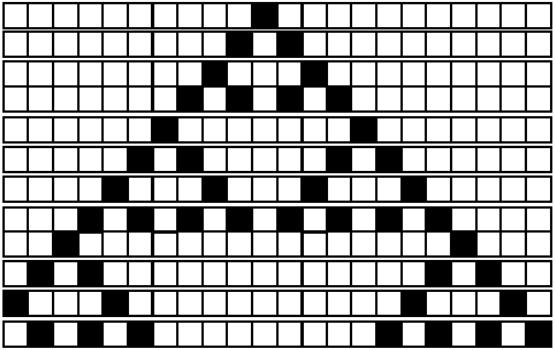

*Figure 4. Automaton no 90*

Given the simplicity of the rules available, and the fact that a priori they are all of the same type (all rules are obtained by putting white or black boxes in the second row of figure 2), we might spontaneously expect all behaviours to be of the same type as those we've just seen, i.e. simple periodic repetitions of straightforward transformations such as a shift or regular patterns. But this is not the case. The surprise comes from the fact that, although on the surface nothing particularly distinguishes certain rules from rule 254 or rule 90, certain other rules lead to behaviours that seems totally random and unpredictable. This is the case, for example, with automaton no. 30 shown in figure 5:

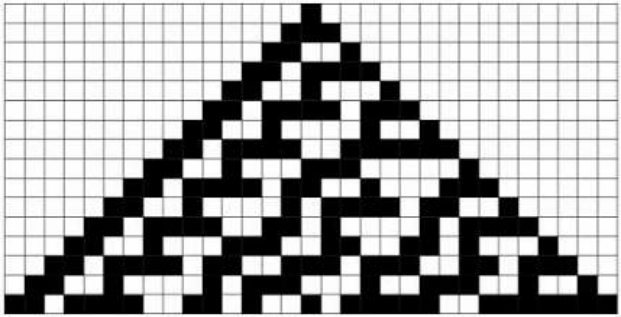

*Figure 5. Automaton no. 30*

This is also the case with automata no. 110 (Figure 6):

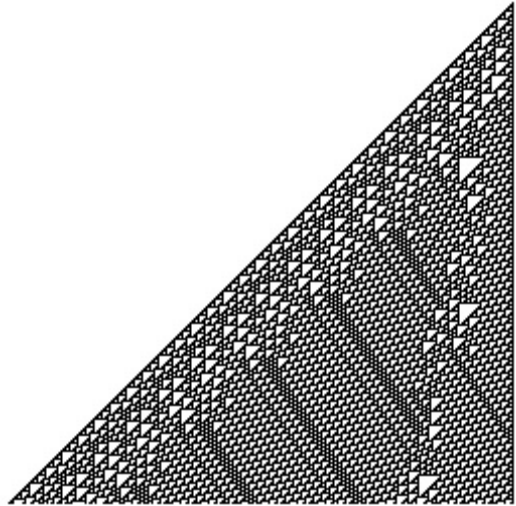

*Figure 6. Automaton no. 110 (large scale)*

The behaviour of these automatons seems to defy any attempt at prediction. More precisely, it seems impossible to have a direct way to directly predict their state at the nth stage without following step by step all the way that leads there. This phenomenon seems to be present in the behaviour of many cellular automata. We will give two other examples.

**2.2 Conway's game of life**

This two-dimensional automaton was invented by British mathematician John Conway in 1970. Cells can be live (black) or dead (white). The rules of evolution are:

- A dead cell with exactly three living neighbours becomes alive (born).
- A living cell with two or three living neighbours remains alive otherwise it dies.

Figure 7 shows the first five steps from the left configuration:

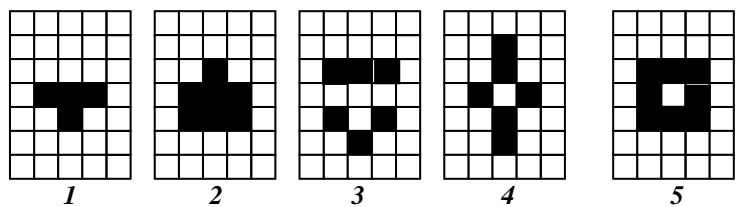

*Figure 7. Example of 5 iterations of the game of life from the configuration 1*

For some initial situations, the following configurations are easily predictable. This is the case, for example, when a configuration is stable and no longer evolves or when a periodic behaviour appears or when a regular movement of cells characterizes the transition from configuration n to configuration (n+1).

Some configurations have special properties. The glider for example (figure 8), moves one square along a diagonal in 4 generations.

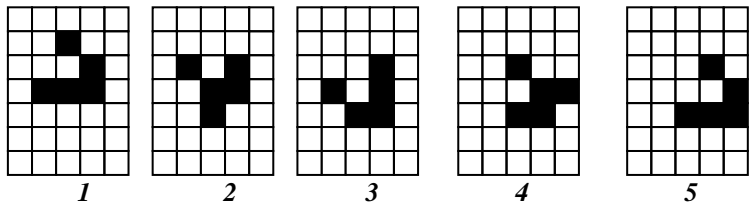

*Figure 8. The glider*

But some initial situations defy any attempt to predict their evolution and the only way to know what they become after for example 1000 steps, is to go through the 999 intermediate steps. Who could predict from the simple view of the configuration of the following figure 9:

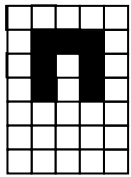

*Figure 9. What will this configuration give?*

that it gives the beautiful clown face of figure 10 after the 110th iteration?

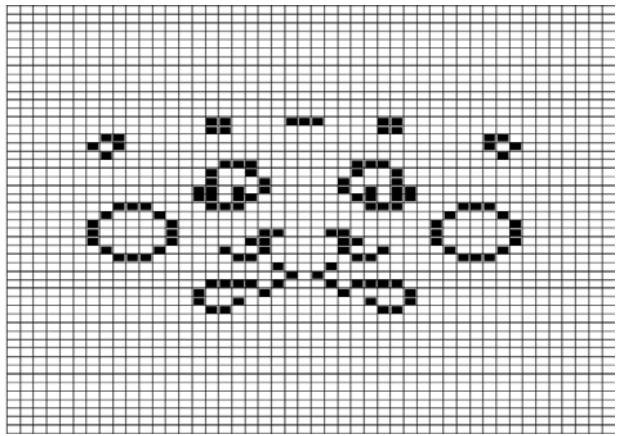

*Figure 10. Clown face*

In the same way who could imagine that the configuration of figure 11, called a glider gun, makes a glider every 30 steps:

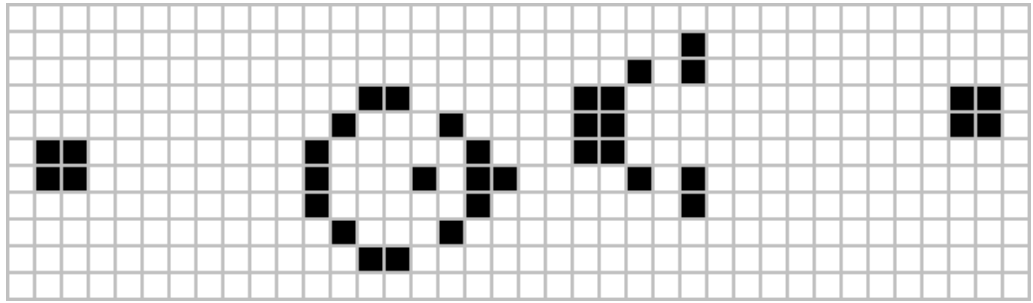

*Figure 11. Glider gun*

The only way to find it out is to run a simulation and see what happens. For the initial configurations of the game of life, there seems to be no other way of knowing how they evolve than by observing them in a simulation. There is no formula for calculating the nth step directly from the initial configuration. It can even be shown that knowing the ultimate fate of an initial configuration is an undecidable problem: there is no general algorithm capable of predicting, when presented with an input configuration, whether it will eventually die out or keep cells indefinitely.

**2.3 Langton's ant**
Another surprising cellular automaton is that of the Langton's ant. Its description is very simple: The ant moves on the squares of a grid on the left, right, top, and bottom.
- If the ant is on a black square, it turns 90° to the right, changes the colour of the square to white and advances one square.
- If the ant is on a white square, it turns 90° to the left, changes the colour of the square to black and advances one square

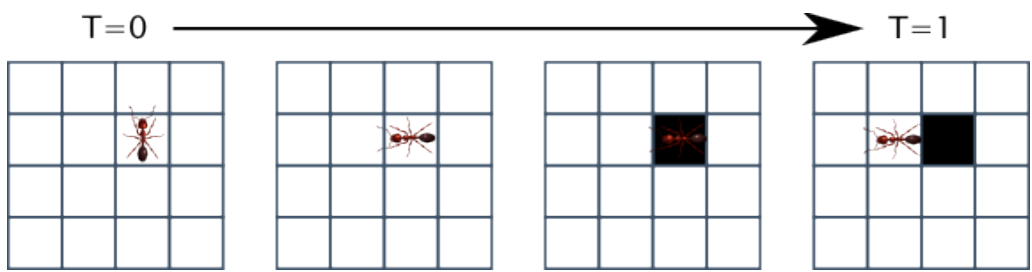

*Figure 12. Departure of the Langton ant from a white square*

After approximately 10,000 iterations during which it has a seemingly random path, the ant goes to infinity by building a straight line formed by a 104-step pattern that repeats itself indefinitely. No one can now prove this result, which remains an observation. Predicting this particular behaviour from the data of the rules of operation yet very simple which presides to its displacement is today impossible and if one asked any mathematician who had never seen explicitly what a Langton's ant does, what is the trajectory of the ant in the long term, it would be totally unable to respond to it unless to simulate step by step the path of the ant.

**3. The computational irreducibility**

**3.1 Intuitive definition**
The examples above provide an intuitive understanding of what is meant by computational irreducibility. For a system that evolves in stages or time steps, it's the impossibility of directly predicting stage n without first having gone through all the previous stages. In a figurative way, we could say: "there is no formula that directly gives step n". But the notion of formula is vague, because the operations that can be used in a formula can be very diverse, and without specifying which operations are authorized, the notion of formula has no precise meaning. It is more accurate to think of computational irreducibility as the fact that the only way to know step n is to first enumerate the (n-1) previous steps, and that there are no shortcuts to "cut to the chase". This was Wolfram's initial definition of the concept[5]. Unfortunately, this simple definition is not satisfactory, as it lacks robustness in the following sense: we would like a system for which it is not possible to predict step n directly, but such that step n can be reached by following (n-1) steps not too different from the (n-1) steps actually followed by the system, to be qualified as computationally irreducible even if the definition we have given is not exactly satisfied. The difficulty then lies in defining what we mean by steps that are not too different. In what follows, we shall see that it is possible to give a precise meaning to this term, leading to a rigorous and robust formal definition.

### 3.2 Formal definition

To give a rigorous and robust formal definition of the concept of computational irreducibility, we need to place ourselves in the context of a specific computational model. There are many computational models (recursive functions, lambda-calculus, Turing machines, RAM machines, etc.) whose equivalence can be shown, i.e. the fact that these models can compute exactly the same functions. Church's thesis states that these functions are exactly the same as computable functions, and that any computable function is computable in one of these models. We will adopt the Turing machine model, as it is perfectly suited to our goal of being able to easily model the path followed during computation, and to reason about computation times to obtain a result. More precisely, we will be working with Turing machines with 3 symbols (0, 1, #), k working ribbons (k ≥ 2) and an output ribbon which will be write-only[6]. The reason for adding the # symbol is the need to be able to explicitly separate intermediate results on the output ribbon[7]. The reason for working with several working ribbons is that calculation times on a single ribbon machine are not the best that can be obtained for a given calculation, but theorems exist showing that optimal times are obtained when several ribbons are allowed[8].

A Turing machine will be said to compute a function f if, when given the input value n, it computes f(n) for all n. We then define a new type of Turing machine that we shall call enumerative Turing machines (E-Turing machines for short). A Turing machine $M_f$ calculating a function f will be called an E-Turing machine for f if, and only if, when the calculation of f(n) is complete, the various values of f(i) for i varying from 1 to n are written in successive order on the output tape (see figure 13).

---

[5] S. Wolfram, «Undecidability and intractability in theoretical physics », *Phys. Rev. Letters*, Vol 54, N 8. 1985.

[6] Here we give only the general principles, omitting many technical details. Interested readers may refer to the following two articles: H. Zwirn, and J.P. Delahaye,, «Unpredictability and Computational Irreducibility», *in* H. Zenil (ed.), I*rreducibility and Computational Equivalence: Wolfram Science 10 Years After the Publication of A New Kind of Science*, Springer, 2013, and H. Zwirn, « Computational Irreducibility and Computational Analogy », *Complex Systems*, 24, 2015, where the theorems justifying the results stated here are also demonstrated.

[7] An identical result could be obtained with two symbols but at the cost of a complication of the coding of the data by what is called "a self-delimited code.

[8] However, it should not be assumed that calculation times decrease in proportion to the number of tapes. A well-known complexity theory theorem shows that the maximum gain between one ribbon and k ribbons (whatever k) is at best quadratic.

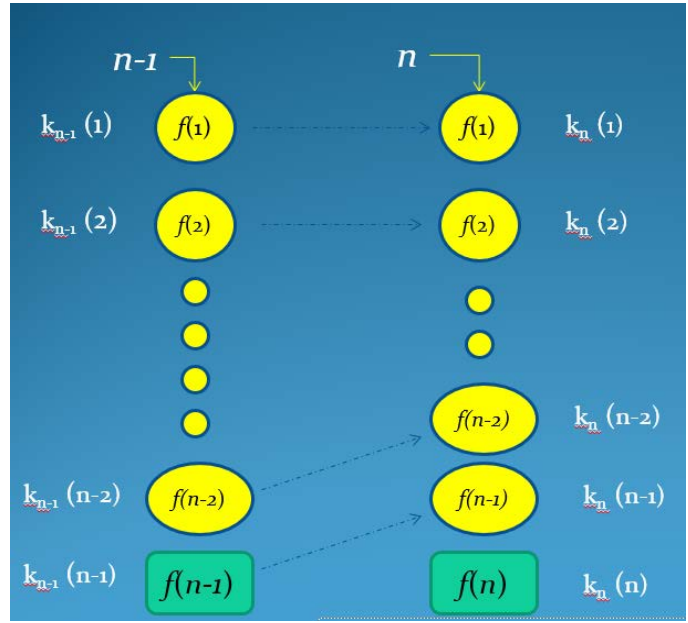

**Figure 13.** *E-Turing machine calculating f(n-1) (left) and f(n) (right)*

The $k_n$ (i) numbers written next to each intermediate result represent the number of steps that the Turing machine performed when it wrote the result in question. So when the machine writes the result f(j) during its calculation of f(n-1), it accomplished $k_{n-1}$(j) steps and when it writes the same result during its calculation of f(n), it accomplished $k_n$ (j) steps. It may happen that $k_{n-1}$ (j) = $k_n$ (j) but it is not mandatory.

We'll need to refer to the computation time it takes a Turing machine to calculate a function f. Complexity theory uses precise notations to express orders of magnitude of computation time in relation to the size of the input data, or more precisely, the way in which this computation time grows in relation to the size of the input data. We will only use the notion $O$ here. To say that a Turing machine computes a function f in time $O(n^2)$ means that the computation time of f(n) will grow approximately as the square of the input data[9].

Clearly, depending on how we go about calculating f(n), we can be more or less efficient. If, for example, unnecessary intermediate calculations are added to the calculation of f(n), the calculation time will increase. On the other hand, it is not obvious that you can always improve the calculation to reduce the time needed. In fact, we shall call efficient Turing machine for f, denoted $M^*_f$, the fastest Turing machine for computing f, a Turing machine such that no other machine computes f faster. Let's denote $T(M^*_f)$ the computation time of such an efficient machine. In the same way, we shall call efficient E-Turing machine for f, denoted $M_f^{eff}$, the fastest E-Turing machine that computes f[10]. There may be several efficient machines that compute the same function. In this case, they all compute at the highest possible speed.

Remember that we said that the difficulty lies in defining what is meant by steps that are not too different. In the context of our calculation model, this comes down to defining what is meant by a calculation path that is not too different from the calculation path followed by the machine calculating f. It is thanks to $T(M^*_f)$ that we will be able to define a distance. To do so, we define the concept of P-approximation of an E-Turing machine for f. This is a Turing machine M such that there exists a function $F(n)=O[T(M^*_f)(n)/n]$ and a Turing machine P such that :

- if given n as input, M computes a result $r_n$ such that if given as input to P, P computes f(n) in a time F(n)
- during the computation of $r_n$ by M, M writes on its output ribbon results $r_i$ (for i ranging from 1 to n-1) such that if $r_i$ is given as input to P, P computes f(i) in a time F(i).

---

[9] It is said that a function k(n) is $O$(p(n)) if and only if there are constants c > 0 and $n_0$ > 0 such as $\forall n > n_0$, |k(n)| ≤ c |p(n)|. It is possible to rigorously justify why this way of expressing calculation times is appropriate.

[10] This definition raises a number of questions that we will leave aside here, if only because of the existence of such machines. We refer the reader again to the two above-mentioned technical articles for a more rigorous presentation.

A P-approximation of an E-Turing machine for f is therefore a machine which, when given n as input, calculates a result $r_n$ which is close to f(n) in the sense that it is possible to calculate f(n) in a very short time from $r_n$, and such that during the calculation of $r_n$, M calculates intermediate results which are also close to the intermediate results f(i) that a true E-Turing machine for f would calculate in the sense we have just specified (see Fig. 14).

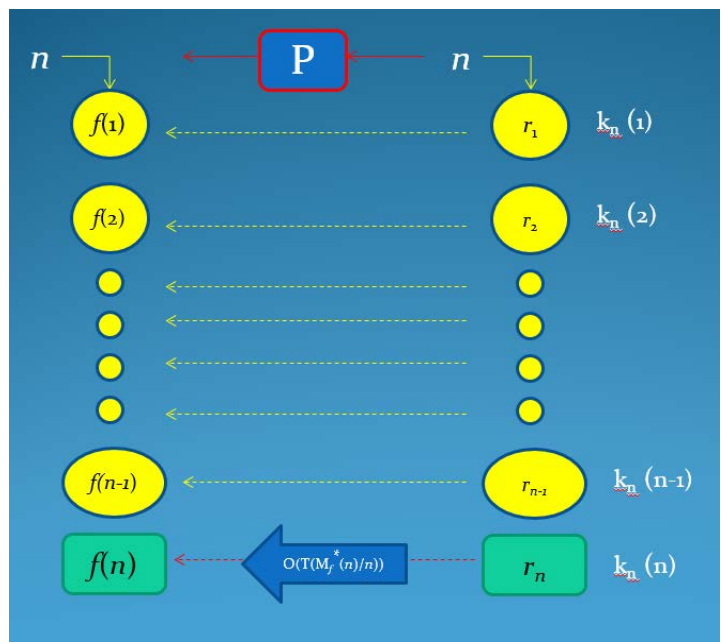

**Figure 14.** *The machine on the right is a P-approximation of the E-Turing machine on the left.*

It is then possible to define the concept of the computation of f based on an approximation. We will say that a calculation of f(n) is based on a P-approximation of a E-Turing machine if f(n) is calculated by first obtaining $r_n$ through the machine which is an approximation then to finish the calculation by calculating f(n) through the P-machine supplied with $r_n$ input (see Figure 15).

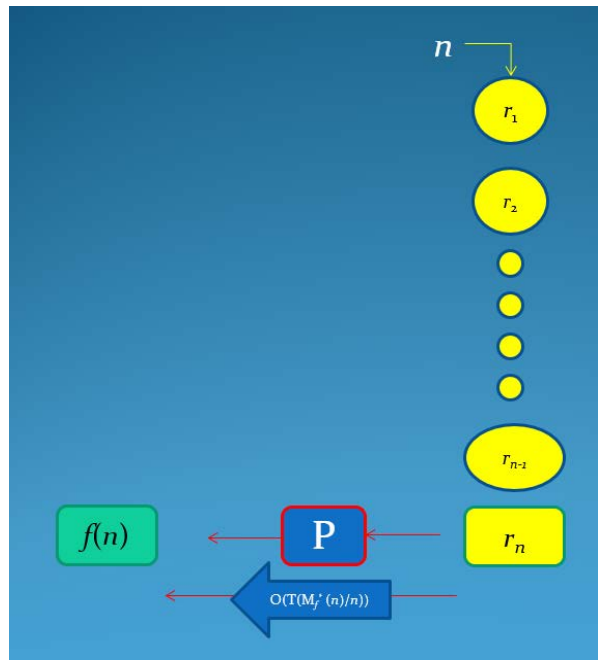

**Figure 15.** *Computation of f(n) based on a P-approximation of a E-Turing machine*

It is at last possible to define a computationally irreducible function:
A function f will be said to be computationally irreducible if and only if for any Turing machine M computing f, there exists a P-approximation of an E-Turing machine for f, M', such that the computation of f by M is based on M'. This definition thus formalizes the notion of a computation path close to the path followed by the machine that computes f and, moreover, expresses the fact that it is impossible to compute f without passing through all these intermediate steps.

It is then possible to prove a number of theorems about computation times, such as the one which states that if a function f is computationally irreducible, no Turing machine can compute this function faster than an efficient E-Turing machine for f.

It remains to be rigorously demonstrated that the behaviour of the Langton ant or the 110 automaton are computationally irreducible with the meaning of the definition given above. Other possible candidates for computational irreducibility include:

1) It is known that the problem of knowing whether an initial configuration of the game of life will be eternal or disappear is undecidable (that is, there is no general algorithm capable of giving the answer when given any configuration in input). Let f(n) the function defined as the number of initial configurations with a number less than or equal to n (in any given configuration numbering) that are still alive after n iterations.

It is obviously not only in the field of cellular automata that candidates for computational irreducibility exist. Here are two other examples:

2) Let f the function defined by: given the irrational number a
- f(1) is the first decimal place of a

- f(2) is the number represented by the following f(1) decimals
- ....
- f(n) is the number represented by the following f(n-1) decimals

3) Let B = {0.1} and B* all the finite strings on B. Let L be a recursive language and consider a numbered list of the words of B* (for example the index in a lexicographic order by increasing lengths). Consider the function f(n) defined by the number of words $w_i$ of B* (for i<n in the chosen numbering) that belong to L.

For different reasons, it seems that these functions are all computationally irreducible. Unfortunately, the rigorous evidence that these functions are really computationally irreducible is currently out of reach and it is therefore an open mathematical problem to formally demonstrate the existence of computationally irreducible functions.

## 4. Emergence explained

To occur, emergent phenomena require the existence of at least two levels (there may of course be more). The first level, often called the micro or local level, is that of the constituents of the system, which are governed by certain laws of behaviour. A second level, called the macro or global level, is where the emergent property appears. This is a level where the basic entities are no longer the constituents themselves, but sets of constituents. Frequently, the emergent property applies only at the global level, and has no meaning at the local level. For example, the property of water of being liquid at room temperature and atmospheric pressure is a property that applies to a set composed of many water molecules, but makes no sense for a single water molecule. It's an emergent property, because it is far from obvious from the individual properties of water molecules, and even less so from the constituents of these molecules, namely hydrogen and oxygen, both of which are gaseous under the same conditions. It is only by observing an assembly of water molecules that we can observe the phenomenon of liquidity. Similarly, the property of the glider gun in the game of life to produce gliders can only be defined in relation to the consideration of a set of cells (a glider) characterized both by its geometry and its behaviour (this set moves while retaining its shape over time). The emergent property of generating gliders can therefore only be defined at the global level, and has no meaning at the level of an individual cell. What qualifies a property as emergent is the fact that even perfect knowledge of the properties of constituents at the local level is not enough to anticipate the appearance of the property at the global level. There is no intuitive and immediate link between knowledge of the local level and the emergent property at the global level: hence the surprise effect often put forward. It is sometimes said that we don't understand what is happening.

Understanding a phenomenon means identifying the rules that govern its behaviour, and being able to mentally follow the steps that, from a known initial state, lead to the final state to be explained, or to the appearance of the phenomenon. If repeated application of the simple rules of a cellular automaton clearly shows that a particular pattern will appear (for example, alternating black and white cells), we can say that we understand the behaviour of this automaton and that it is not emergent. If, on the other hand, the repeated application of the rules is not mentally possible (e.g. for the automaton obeying rule 110), then we call the behaviour emergent and feel that we do not understand it. On reflection, however, it seems that the two situations differ only in degree, and are not really different in nature. Part of "understanding" is being able to mentally string together the steps in a process leading from an initial state to a final state. When these steps are too numerous or too intricate for this to be possible, we have to resort to computer simulation, and the sensation of understanding fades away. We can therefore defend the idea that comprehension is in part the possibility of mental simulation. When the mental simulation is simple and the result is easily anticipated, there is no emergence.

There are now different levels of complexity in the difficulty of reproducing emerging phenomena from the local level. As we said in the introduction, emergent phenomena can occur in more or less complex environments, making it more or less easy to identify them as such. In some cases, we can easily reproduce the emergent phenomenon a posteriori by simulation, based on the properties of the constituents. This is, of course, the case with the examples we have given of cellular automata, since the process for bringing out the emergent property is precisely the computer simulation of the behaviour of the constituents. In this case, we have no immediate intuitive understanding of how the basic rules produce the emergent phenomenon, but we know very easily how to chain these basic rules (this is the very principle of simulation) to make it emerge. However, in the case of real physical phenomena occurring in a complex environment, it may be difficult or even impossible to reproduce the emergent phenomenon computationally on the basis of the laws governing its constituents. This is the case, for example, in physical chemistry, where a calculation of the properties of liquid water would have to start with the properties of the constituent hydrogen and oxygen atoms and the laws of electrodynamics and quantum mechanics (known as an ab initio calculation). Although great progress has been made in this field, to my knowledge, ab initio calculation of many water properties is still out of reach. But with the rapid growth in computer power, it may well be possible in the near future. We will then be in the same situation with water than with cellular automata: it will be possible to calculate emergent properties, but they will continue to elude direct intuition. These examples concern systems for which the local rules are known. The difficulty with water is mainly a practical difficulty in carrying out the calculations, whereas they are easy in the case of cellular automata. On the other hand, this calculation is totally out of reach for emerging biological phenomena such as life or consciousness. Apart from the practical difficulty of carrying out the calculations, which are undoubtedly immeasurably more complex than for water, we know virtually nothing about the underlying rules that would have to be simulated. The enormous progress made in biology and neuroscience in recent years is still a long way from giving us the key to an intimate understanding of these phenomena. This is already true for life, but understanding the phenomenon of consciousness seems even further away. Understanding consciousness is undoubtedly the most difficult problem facing science all disciplines merged. But despite these differences in degree, the very essence of the emergence of all these phenomena remains the same: there is no way of anticipating the result they will produce simply by looking at the basic rules, because there is no algorithm for skipping steps to arrive directly at the result. So, to know the final step, we need to go through all the intermediate stages, and our brains are not powerful enough to intuitively tell us what the result will be simply by knowing the starting point and the rules we have followed.

The explanation we arrive at is therefore a hybrid one. It includes an objective part, which is the computational irreducibility of the processes that make up the dynamics of the system: there is no algorithm that calculates the final result without first calculating the intermediate results (or results close to them). This is the essence of emergent phenomena. But there is also a subjective part, due to our inability to immediately foresee the outcome of a complicated sequence of many successive steps. It is this inability that is the cause of the surprise we feel when we observe the result. As a result, for us, objective emergence is necessarily subjective too. But beings with far greater mental computational capacities than ours, for whom the simulation of a few thousand steps in a cellular automaton would be almost instantaneous, would probably not feel the same surprise at the behaviour of Langton's ant or the configurations of the game of life. The objective emergence of certain phenomena might not be associated for them with subjective emergence. The fact remains, however, that if their capacities are not infinite, they would also have limits, and thus, astonishment would come back for any configuration whose emergent effect lies beyond these limits. Surprise in the face of an emergent phenomenon would therefore only be a matter of degree, while computational irreducibility would always be present.